# Comparison of the upper atmosphere temperature obtained by ground-based and satellite instruments

Saunkin A.V., Vasilyev R.V., Artamonov M.F.

Institute of Solar-Terrestrial Physics, Siberian Branch of the Russian Academy of Sciences, Irkutsk, Russia, saunkin@irk.iszf.ru
Irkutsk State University, Irkutsk, Russia

**Abstract.** An approach is demonstrated for comparing the temperature of the upper atmosphere obtained by ground-based and satellite methods. A method for calibrating ground-based instruments (Fabry-Perot interferometer) based on the data obtained and determining the exact altitude at which the temperature was measured by ground-based means is proposed. Based on the assumption that the temperatures are connected in a linear fashion, the coefficients of the temperatures connection are determined by the least squares method. We use information on the temperature of the upper atmosphere for 2017-2018 obtained by the SABER TIMED and MLS Aura satellite instruments, and by the Fabry-Perot interferometer located at the Tory geophysical observatory of the ISTP of the SB RAS observing the 557.7 nm oxygen line, the emission height is ~ 100 km. The scale factor $a \approx 0.99$ and the displacement coefficient $b \approx 120K$ were obtained for the temperature data using an interferometer at a height of $h = 92$ km for which the seasonal variation of the temperatures of the interferometer most closely corresponds to the seasonal variation of the temperatures obtained by SABER TIMED. The coefficients can then be used as calibration values for the interferometer.

*Atmosphere of Earth, satellite data, satellite data processing, Python, Matplotlib, Cartopy, NetCDF4, H5py, Numpy.*

**Introduction**

To check the plausibility of the output data, experimental installations, it is required to compare the results obtained with other sources of information to identify inconsistencies, possible errors, for their subsequent elimination. The temperature of the upper atmosphere is usually determined by remote sensing methods. Direct measurements are rare both in time and space. Satellite and ground-based measurements are more common. The physical foundations of methods for observing temperature by satellite and ground-based means are different; therefore, it is often necessary to compare the data obtained. For example, in [1], a comparison was made of the temperature obtained by satellite and ground-based instruments at altitudes of 84 and 85 km for geographic regions close to Irkutsk and Zvenigorod, where a negative bias of satellite data compared with ground-based data was revealed. The authors of [2] use the satellite data SABER TIMED [3] and MLS Aura [4] to calculate the equivalent temperatures of the hydroxyl layer for comparison with the values measured from the OH (6-2) emission lines observed by the ground-based spectrometer located at Davis station in Antarctica.

In this paper, we compare the data on the temperature of the upper atmosphere obtained using SABER TIMED and MLS Aura and temperature data obtained using the Fabry-Perot interferometer (IFP) located at the geophysical observatory of ISTP SB RAS "Tory" (51.81 N, 103.07 c. d.). The IFP is capable of observing the temperature of the upper atmosphere at different altitude levels [5]. For the studies carried out in this work, we chose the data obtained by the FPI from the observation of the oxygen line at 557.7 nm. The disadvantage of FPI is the absence of a calibration light source in the optical range of about 577.7 nm. It is at this wavelength that atomic oxygen radiates at an altitude of about 100 km in a layer about 5-10 km thick. Thus, the lack of information about the hardware function of the FPI can lead to a distortion of the temperature observed along the green oxygen line. To eliminate this ambiguity, this paper attempts to compare the temperatures obtained from independent sources and to calibrate the ground-based facility using satellite data.

The initial satellite data on the temperature of the Earth's upper atmosphere were obtained from the following databases:

1) The site of the MLS Aura instrument (https://mls.jpl.nasa.gov/index-eos-mls.php) conducting limb observations of the atmosphere to obtain the parameters of the altitude distribution.
2) The SABER TIMED instrument website (http://saber.gats-inc.com/data.php) also provides limb atmospheric observations and provides demo distributions of atmospheric parameters.

Satellite data posted on these resources are stored in various formats (HDF4, HDF5, NetCDF, etc. (https://annefou.github.io/metos_python/02-formats). The information presented in the files represents altitude profiles of a set of parameters, including temperature, therefore, to obtain specific physical and chemical parameters of the atmosphere in given coordinates above the Earth's surface, it was necessary to develop a software package with appropriate algorithms using libraries for "Python" [6]. Data visualization was carried out using the packages " Matplotlib "[7] and" Cartopy "[8], reading of various data formats was carried out using the packages" H5py "[9]," NetCDF4 "[10], mathematical processing of the obtained information was carried out using the package" Numpy "[11].

Fig. 1 shows the results of visualization of the distribution of measurement points of temperature profiles for various satellites above the geophysical observatory "Tory" during 2017-2018. The area of satellite measurements corresponds approximately to the area of temperature measurements carried out by the FPI.

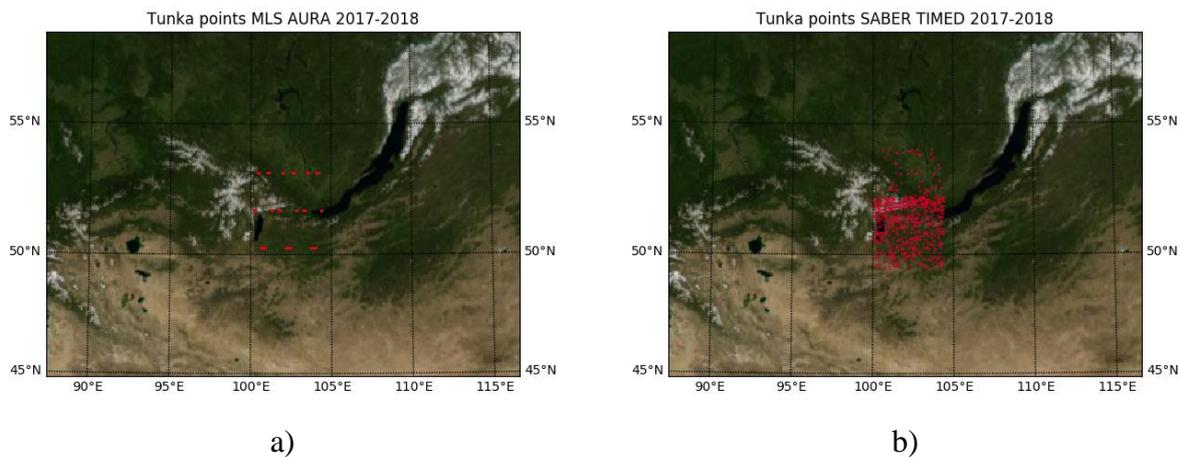

a)                                                                 b)

Fig1. Accumulated points of measurements in a certain area from different satellite instruments: a) MLS Aura, b) SABER TIMED

Visual assessment of maps in Fig. 1 says that the number of measurements per time of the MLS Aura is less than SABER TIMED. The reason is that MLS Aura has a sun-synchronous orbit, so the satellite flies along the same trajectory each time and ends up over the same measurement points at the same time of day. The total number of measurements for the specified period of time over the selected area is approximately the same: for SABER TIMED 727, for MLS Aura 772.

In fig. 2 shows a comparison of data from two satellites at an altitude of about 90 kilometers on the considered area and in chused period of time. For each measurement carried out by the SABER TIMED satellite, the closest MLS Aura measurement points in time were selected. An estimate of the time intervals between measurements by two satellite instruments is shown in Fig. 3, here, in the form of a statistical distribution (histogram), the characteristic scatter of the difference between the MLS Aura and SABER TIMED measurements is presented; the time interval between successive measurements can reach two days, since the satellites have different trajectories and time of flight over a certain area. Nevertheless, the temperature data obtained from the SABER TIMED and MLS Aura satellites for the selected area show an identical seasonal variation over the selected time interval.

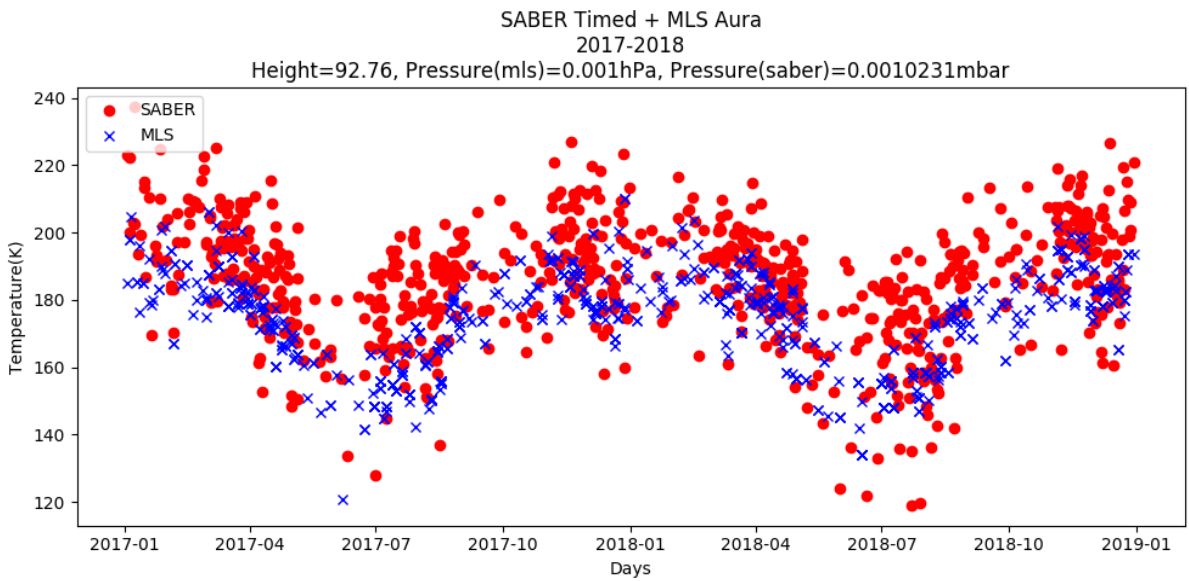

Fig.2 Temperature variation obtained by satellite instruments over the considered area for the considered period of time at an altitude of ~ 90 km

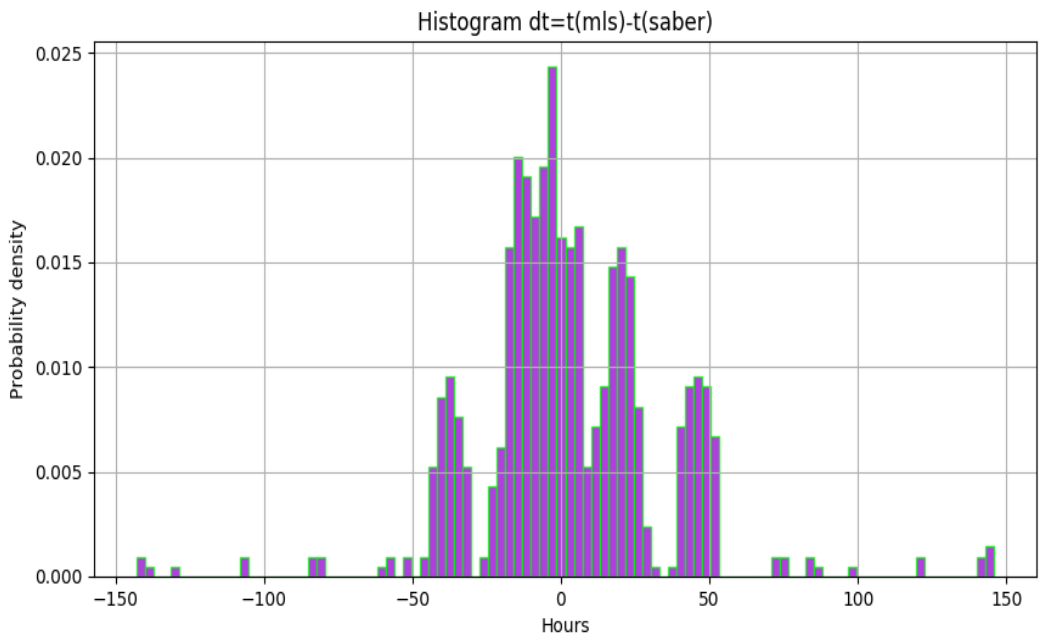

Fig. 3 Histogram of time Δt – the difference between the time of measurements of satellite means Fig. 2.

In the picture fig. 4 shows the distribution of satellite temperature measurements for an altitude of about 90 km by time of day. For MLS Aura, measurements, as mentioned above, take place at a certain fixed time - at ~ 6: 30 and ~ 19: 30 UT, and for SABER TIMED they are distributed over the day, and the measurements are mainly taken in the evening.

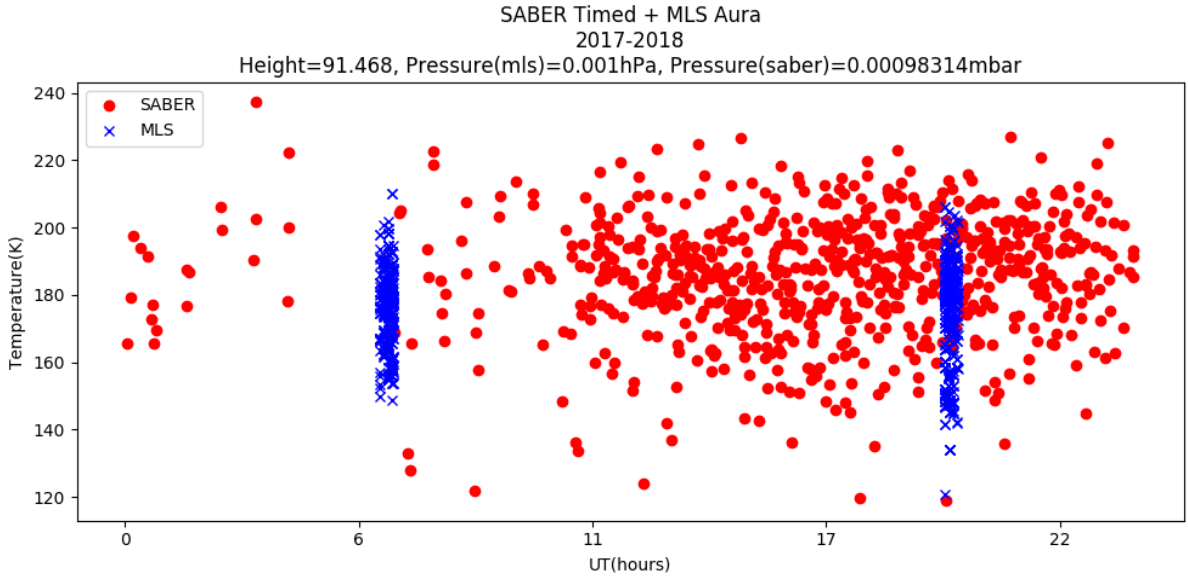

Fig. 4 Daily accumulation of temperature measurements for two years

SABER TIMED has the largest scatter of measurements over the time of day and a more uniform coverage of the territory, in contrast to MLS Aura and will be used in the future as more promising for comparison with the ground-based Fabry-Perot interferometer, since the Fabry-Perot data may be obtained unevenly at night due to background lighting conditions (moonlight).

Temperature data at an altitude of about 100 km obtained with a Fabry-Perot interferometer, as mentioned above, may contain some error. Let us assume that the temperature obtained by the satellite and ground-based means is linearly related:

$$T_i^h = aT_i + b \quad (1)$$

where $T_i$ is the IFP temperature, $T_i^h$ is the SABER TIMED temperature obtained at the height h, a and b - the required coefficients. The coefficients of this linear relationship can be determined using the method of least squares. The task is to find the minimum value of the sum of squares of the error function by varying coefficients a and b:

$$S(a,b,h) = \sum_{i=1}^{n}(aT_i + b - T_i^h)^2 \quad (2)$$

To find the minimum, we find the partial derivatives of the function S (a, b) with respect to variables a and b, we equate these derivatives to zero, solve the resulting system, and then the sought coefficients are expressed as:

$$a = \frac{\sum_{i=1}^{n} T_i T_i^h - nb \sum_{i=1}^{n} T_i}{n \sum_{i=1}^{n} T_i^2} \quad (3), \qquad b = \frac{\sum_{i=1}^{n} T_i^h - na \sum_{i=1}^{n} T_i}{n} \quad (4)$$

Since the exact height of illumination in this work is assumed to be unknown, it is necessary to perform this operation for some set of heights, on which SABER TIMED monitors. Then it is necessary to correct the temperature obtained by the FPI for a linear relationship with the SABER TIMED data for each altitude level, and then this corrected temperature can be analyzed using the Pearson correlation coefficient or the behavior (2) with altitude for the already obtained parameters a and b can be analyzed. In the first case, the height of the maximum of the correlation coefficient, and in the second case, the height of the minimum (2) will indicate the height where the temperature is actually measured by two means simultaneously.

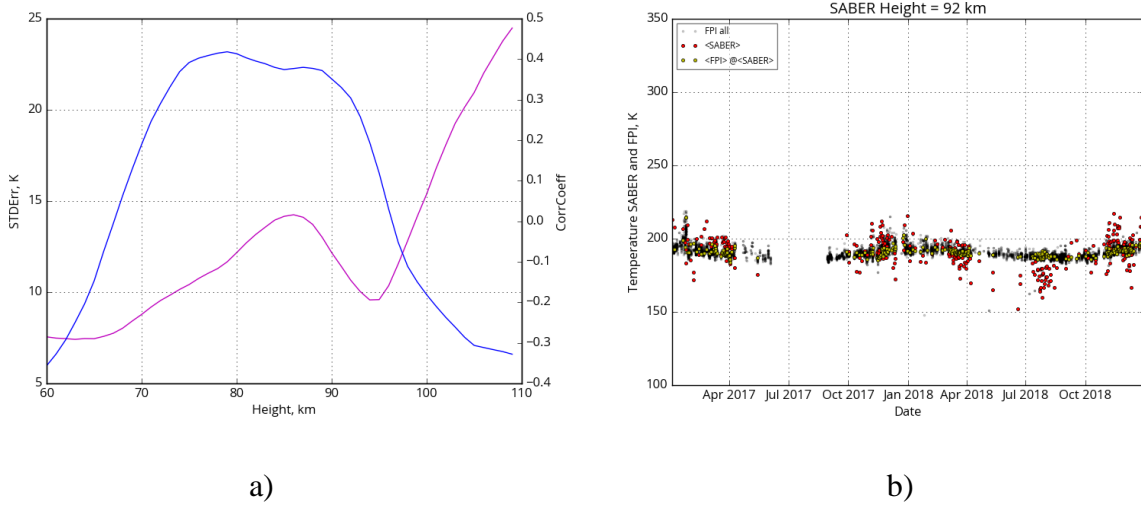

a)                                  b)

Fig. 5 Results of SABER TIMED and IFP visualizations: a) error function S (a, b) as "STDErr" purple color and Pearson correlation coefficient "CorrCoeff" blue color in relation to height, b) seasonal temperature behavior: red - temperatures obtained by SABER TIMED, dark - temperatures obtained by FPI, yellow - Fabry Perot temperatures averaged over a two-hour interval relative to the SABER TIMED measurement time.

Figure 5, a) shows the behavior of these parameters with height. The behavior of the correlation coefficient with altitude has two local maxima, at heights of 79 and 89 km. The error function has one local minimum at an altitude of ~ 95 km, a scale factor a ≈ 0.05 and a bias b ≈ 160K for this height. However, if we analyze the results of the behavior of the corrected FPI temperatures and satellite temperatures in (Fig. 5, b), it can be seen that the data do not coincide very well, especially their dispersion. A possible cause could be a wrong approach to minimization, since the scale factor and bias, the coefficients "a" and "b" respectively, are minimized together. In addition, the data obtained by the FPI have a seasonal temperature variation (about 20K) about half that of the SABER TIMED (about 40K).

To minimize the experimental data about the temperature, taking into account the bias and the scale factor separately, the error function was divided into two parts, separately to minimize the bias (coefficient b):

$$a = 0 : S(b,h) = \sum_{i=1}^{n} \left(b - T_i^{\,h}\right)^2 \quad (5)$$

at $\frac{dS}{db} = 0$ we obtain:

$$b = \frac{\sum_{i=1}^{n} T_i^{\,h}}{n} \quad (6)$$

where b is the average value of $T_i^{\,h}$, and separately to minimize the scale factor (coefficient a):

$$b = 0 : S(a,h) = \sum_{i=1}^{n} \left(aT_i - T_i^{\,h}\right)^2 \quad (7)$$

for $\frac{dS}{da} = 0$ we get:

$$a = \frac{\sum_{i=1}^{n} T_i T_i^{\,h}}{n \sum_{i=1}^{n} T_i^{\,2}} \quad (8)$$

The result of a two-stage minimization, in which the bias was initially determined, and then the scale factor is shown in Fig. 6. Despite the fact that (Fig. 6, a) there are two local minima in both considered characteristics, the first, main local minimum is located too low from the height at which oxygen glow occurs, therefore we will consider the second local minimum. For an altitude of 89 km, where the second maximum of the Pearson correlation coefficient is located, it is also not physically justified to choose the scale factor and the shift, since it is well known that the glow of the hydroxyl layer is realized at this height, and a significant amount of previously performed studies separates the emission height of the hydroxyl from the emission height of atomic oxygen (for example, [12]). Therefore, it is most advisable to choose for determining the glow height (2). It is for an altitude of 92 km, where the second local minimum (2) is located, that the displacement a ≈ 0.99 and the scale factor b ≈ 120K were chosen. The seasonal variation of the FPI corrected for these temperature coefficients is shown in (Fig. 6, b). If we compare (Fig. 5, b) and (Fig. 6, b), then it can be noted that in the second figure the data are more consistent with each other and there is a coinciding seasonal variation of temperatures, that is, in the second approach to minimization, we obtained correct coefficients, in contrast to first.

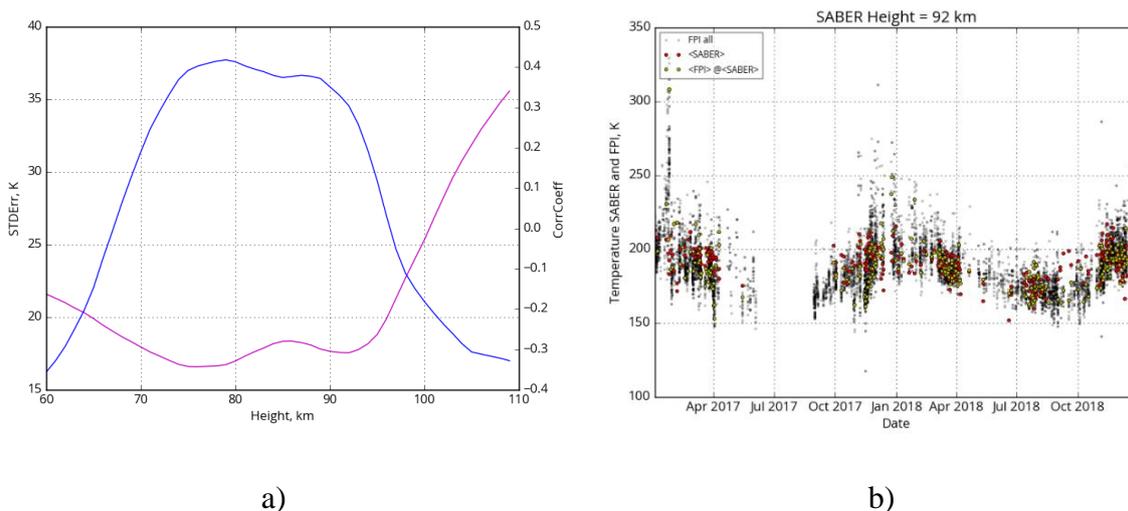

a) b)

Fig. 6 Results of visualization of SABER TIMED and FPI temperatures after separation of the method of least squares formula: a) error function S (a, b) as "STDErr" purple color and Pearson correlation coefficient "CorrCoeff" blue color in relation to height, b) seasonal behavior of temperatures: red - temperatures obtained by SABER TIMED, dark - temperatures obtained by FPI, yellow - Fabry Perot temperatures, averaged over a two-hour interval relative to the time of SABER TIMED measurements.

**Conclusion**

The satellite data of two MLS Aura instruments, SABER TIMED were considered, and visualization and comparison were performed to identify more informative in time and space data. When comparing, it was found that the SABER TIMED data are the most suitable both in terms of the spatial component and the timing of measurements (mainly in the night time) for further comparison of satellite data with a ground-based instrument - the Fabry-Perot interferometer. As a result of the comparison, the assumption about the presence of an error was confirmed. Therefore, when searching for a linear relationship, the least squares method was used to obtain the coefficients a and b, the scale factor and the bias, respectively. In the course of solving the problem, two approaches were used. In the first (standard) approach, the data do not coincide very well, and the seasonal temperature variation is poorly expressed. In the second, when analyzing the a and b coefficients separately, the data obtained are more consistent with each other, and a coinciding seasonal temperature variation is observed. As a result of solving the problem from the second approach, the scale factor coefficients a $\approx$ 0.99 and the displacement b $\approx$ 120K were obtained, which can then be used to correct the data on the temperature of the upper atmosphere at an altitude of about 100 km obtained by the Fabry-Perot ground instrument.

The direct comparison of the temperatures presented here is the zero-level approximation for rough calibration of the Fabry-Perot interferometer temperature dataset. Actually one should consider the details of 557.7 nm intensity height profile in such comparisons because the observed thermal Doppler widening of the line should be the linear combination of widenings from different heights with weights equal to the corresponding intensity of the airglow. The temperature gradients near the mesopause level can be very sharp so the comparison of measured on the ground level temperature using 557.7 nm airglow with the satellite temperature data should take in to account both the intensity and temperature profiles.

**Acknowledgments**

2020, the acquisition and processing of satellite data, as well as their comparison with data from ground-based instruments was carried out with the financial support of a grant from the Russian Science Foundation (project No. -17-00042).